\RequirePackage{tikz}
\usetikzlibrary{decorations.pathreplacing,calligraphy,backgrounds}

\documentclass[sn-mathphys]{sn-jnl}

\usepackage{float}
\restylefloat{table}

\jyear{2021}%

\theoremstyle{thmstyleone}%
\theoremstyle{thmstyletwo}%

\theoremstyle{thmstylethree}%

\newcommand{\gf}{\check{g}}
\newcommand{\gp}{\hat{g}}
\newcommand{\tauf}{\check{\tau}}
\newcommand{\taup}{\hat{\tau}}
\newcommand{\mmf}{\check{m}}
\newcommand{\mmp}{\hat{m}}
\newcommand{\Mf}{\check{M}}
\newcommand{\Mp}{\hat{M}}

\newcommand{\op}{\hat{\Omega}}
\newcommand{\of}{\check{\Omega}}
\newcommand{\Lambdap}{\hat{\Lambda}}
\newcommand{\Lambdaf}{\check{\Lambda}}
\newcommand{\lambdap}{\hat{\lambda}}
\newcommand{\lambdaf}{\check{\lambda}}
\newcommand{\Rp}{\hat{R}}
\newcommand{\Rf}{\check{R}}
\newcommand{\Gp}{\hat{G}}
\newcommand{\Gf}{\check{G}}
\newcommand{\rhop}{\hat{\rho}}
\newcommand{\rhof}{\check{\rho}}
\newcommand{\pp}{\hat{p}}
\newcommand{\pf}{\check{p}}

\newcommand{\squaref}{\check{\square}}
\newcommand{\Tp}{\hat{T}}
\newcommand{\Tf}{\check{T}}
\newcommand{\ap}{\hat{a}}
\newcommand{\af}{\check{a}}
\newcommand{\ddf}{\check{\nabla}}

\newcommand{\ddp}{\hat{\nabla}}

\newcommand{\Uf}{\check{U}}
\newcommand{\omegap}{\hat{\omega}}
\newcommand{\omegaf}{\check{\omega}}

\begin{document}

\title[Toward fixing a framework for conformal cyclic cosmology]{Toward fixing a framework for conformal cyclic cosmology}

\author[1]{\fnm{Oliver} \sur{Markwell}}\email{otm16@uclive.ac.nz}
\equalcont{These authors contributed equally to this work.}

\author*[1]{\fnm{Chris} \sur{Stevens}}\email{chris.stevens@canterbury.ac.nz}
\equalcont{These authors contributed equally to this work.}

\affil*[1]{\orgdiv{Mathematics and Statistics}, \orgname{University of Canterbury}, \orgaddress{\street{Science Road}, \city{Christchurch}, \postcode{8140}, \state{Canterbury}, \country{New Zealand}}\newline\newline\small{\emph{Dedicated to Peter Norman Stevens 1946-2022, who inspired}}}


\abstract{Conformal Cyclic Cosmology (CCC) is a cyclic model of the universe put forward by Sir Roger Penrose \cite{penrose2010cycles}. A conformal invariance assumption in the neighbourhood of the crossover region between cycles (which Penrose calls \emph{aeons}) allows successive space-times to be related by a conformal rescaling. A major open problem is how to choose the conformal factor in a unique way, and is a fundamental hurdle to further study. Proposals have been put forward by Newman \cite{newman2014fundamental}, Tod \cite{tod2015equations} and Nurowski \cite{nurowski2021poincare}, but they disagree in one way or another with Penrose's original assumptions as well as each other. In this paper we compare these different models in detail and rule out certain choices for the conformal factor that have been put forward by Penrose. We extend the results of Newman and fix inconsistencies that arose in his calculations. A new class of solutions are put forward which agree with Penrose's assumptions exactly so long as a certain additional relation is satisfied.}




\maketitle

\section{Introduction}\label{sec:intro}
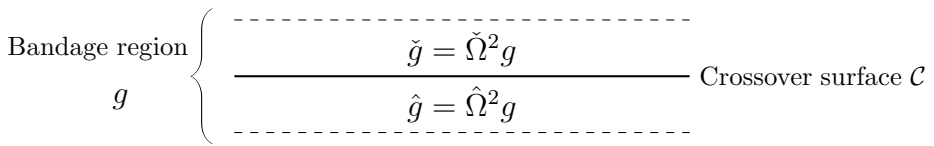
\begin{figure}[h]
    \centering
    \begin{tikzpicture}[scale=0.75]
    \draw[dashed](-4,1) -- (4,1);
    \draw[black,thick](-4,0) --(0,0)node[anchor = north]{\large$\hat{g} = \op^2g$}node[anchor = south]{\large$\check{g} = \of^2g$} -- (4,0)node[anchor = west]{Crossover surface $\mathcal{C}$};
    \draw[dashed](-4,-1) -- (4,-1);
    \draw [decorate, 
    decoration = {calligraphic brace, raise = 4pt, amplitude = 8pt}] (-4.2,-1.2) --  (-4.2,1.2);
    \node at (-6.4, 0.5) {Bandage region};
    \node at (-6, -0.4) {\large$g$};
    \end{tikzpicture}
    \caption{The conformally invariant neighbourhood of the crossover. A hat denotes quantities in the previous aeon while a check denotes quantities in the future aeon.}
    \label{fig:crossover}
\end{figure}

In 2010 Sir Roger Penrose released the book "Cycles of Time" \cite{penrose2010cycles}, a radical new theory of cosmology. The main idea is that the remote future of an expanding universe can be joined to the big bang of another. This possibility exists since a large class of space-times admit both a conformal rescaling of the big bang to a smooth space-like 3-surface \cite{paul2002isotropic, tod2003isotropic} and a conformal boundary in the asymptotic future that is also a space-like 3-surface \cite{hawking1973large}. Penrose dubs each cycle an \emph{aeon}, and postulates that there exists a \emph{bandage region} whereby these two space-times are determined solely by their lightcone structure, i.e. there is no rest-mass and are thus related by a conformal rescaling, see Fig. \ref{fig:crossover}. This cycle of aeons automatically satisfies the strong form of the Weyl curvature hypothesis at the big bang \cite{tod2010penrose} since $C_{abc}{}^d$ is conformally invariant and $C_{abc}{}^d\rightarrow0$ in the asymptotic future.

Whilst this theory requires controversial physical assumptions, such as the asymptotic decay of rest-mass and information loss inside black holes, we do not address them here and instead focus on the mathematical issues surrounding the theory.

In our discussions of CCC, we will concentrate on two consecutive aeons and call them the \emph{past aeon} and \emph{future aeon} to avoid ambiguities that arise by refering to one of them as the 'present'. Quantities in the previous and future aeons are represented by hats $\hat{}$ and checks $\check{}$ respectively, mimicking lightcones. In the conformally invariant bandage region, the metric of each aeon can be related to a \emph{bridging metric} $g$ via
\begin{equation}\label{eq:metricdefns}
    \gp_{ab} = \op^2g_{ab},\qquad
    \gf_{ab} = \of^2g_{ab}.
\end{equation}
In order to make a connection between aeons, Penrose put forward the \emph{Reciprocal Hypothesis} \cite{penrose2010cycles}, 
\begin{equation}\label{eq:reciphyp}
    \op\of = -1.
\end{equation}
This is a very simple way to satisfy $\text{d}\op\neq0$, a requirement for an asymptotically simple space-time, and to match $\op\rightarrow0$ and $\of\rightarrow+\infty$ as $\mathcal{C}$ is approached from the appropriate directions. This will be assumed for the entirety of this paper. Putting Eqs \eqref{eq:metricdefns} and \eqref{eq:reciphyp} together yields the relationship between past and future aeons
\begin{equation}\label{eq:gfTogp}
    \gf_{ab} = \op^{-4}\gp_{ab}.
\end{equation}
Hence, once $\op$ is known, so is the future aeon. The natural question to ask at this point is how can $\op$ be determined? The conformal rescaling of the previous aeon's scalar curvature $\Rp$ to the bridging metric's scalar curvature $R$ is
\begin{equation}\label{eq:RicciTrafo}
    (\Box + \frac{1}{6}R)\op 
    = \frac{1}{6}\Rp\op^3 \iff (\hat{\Box} 
    + \frac{1}{6}\Rp)(\op^{-1}) 
    = \frac{1}{6}R\op^{-3},
\end{equation}
written with respect to different connections for use later. Penrose proposes this must be satisfied with the restriction $\Rp=4\lambdap=R$, where $\lambdap$ is the cosmological constant. This then yields a second order hyperbolic equation for $\op$, determined solely by quantities in the past aeon, called the \emph{Phantom Field Equation}, which fixes the conformal factor up to two pieces of initial data on $\mathcal{C}$. Newman \cite{newman2014fundamental}, Tod \cite{tod2015equations} and Nurowski \cite{nurowski2021poincare} all differ from Penrose by not enforcing that $R=\Rp$ but rather that $R$ is possibly some other constant. There are multiple options in choosing the initial data \cite{tod2021conformal} that have been put forward. However, a unique, physically motivated method to determine $\op$ remains to be seen.

The above yields a rough mechanism for generating $\gf_{ab}$ by solving for $\op$ given $\gp_{ab}$, but what does the Einstein equation in the future aeon tell us? The Einstein equation of the previous aeon in the bandage region is $\Gp_{ab} = \kappa \Tp_{ab}$, where the energy-momentum $\Tp_{ab}$ must be divergence-free, trace-free (zero rest-mass) and can be taken to be a conformal density of weight $-2$ to preserve the divergence-free condition through $\mathcal{C}$. Then this propagates through $\mathcal{C}$ to yield
\begin{equation}
    \Tf_{ab} = \op^4\Tp_{ab},
\end{equation}
Then the Einstein equation in the future aeon is \cite{tod2015equations}
\begin{align}
    \Gf_{ab} + \lambdap \gp_{ab}
    &= -\of^4\Tp_{ab} 
    + 4\frac{\ddf_a\ddf_b\of}{\of} 
    + \frac{\ddf_a\of\ddf_b\of }{\of^2}
    + \Big{(}8\frac{\ddf_c\ddf^c\of}{\of^2} - 4\frac{\squaref\of}{\of}\Big{)}\gf_{ab} \nonumber \\
    &= -\Uf_{ab},
\end{align}
where $\Uf_{ab}$ is the total energy-momentum of the future aeon. It is clear that this has a trace term in general and thus matter can appear in the future aeon. Penrose postulates this corresponds to dark matter. Further, he decomposes $\Uf_{ab}$ into three distinct parts that are conserved individually,
\begin{equation}
    \Uf_{ab} = \Tf_{ab} + \Tf_{ab}[\op] + \check{W}_{ab}.
\end{equation}
Here, $\Tf_{ab}$ is the contribution coming from $\Tp_{ab}$, the energy-momentum content of the previous aeon. $\Tf_{ab}[\op]$ is the contribution of the scalar field $\op$ that is also called the new-improved energy-momentum tensor \cite{penrose1984spinors2} and is trace-free. $\check{W}_{ab}$ is the portion containing the trace-term and the symmetric part is taken so that the Einstein equations in the future aeon are satisfied. The trace-free part has yet to obtain a physical interpretation but Penrose associates the trace term with dark matter.

So far, three similar sets of solutions exist in the literature that satisfy this cyclic model, with variations of Eq. \eqref{eq:RicciTrafo} being used to determine $\op$ \cite{tod2015equations,newman2014fundamental,nurowski2021poincare}. They all consider a radiation fluid FLRW universe on each side of the crossover, given in co-moving and conformal time coordinates respectively as
\begin{align}\label{eq:FLRW}
    \text{d}s^2 
    &= \text{d}t^2 - a(t)^2\Big{(}\frac{1}{1 - k r^2}\text{d}r^2 + r^2\text{d}\sigma^2\Big{)} \nonumber \\
    &= a(\tau)^2(\text{d}\tau^2 - \frac{1}{1 - k r^2}\text{d}r^2 - r^2\text{d}\sigma^2)
    = a(\tau)^2\tilde{g},
\end{align}
where $\text{d}\sigma^2 = \text{d}\theta^2 + \sin^2\theta\,\text{d}\phi^2$ is the line element of the unit 2-sphere and $k=-1,0,1$ determines the spatial curvature. This necessarily means that the trace of the induced energy-momentum tensor that potentially corresponds to dark matter does not make an appearance. The energy momentum tensor of the previous aeon is
\begin{equation}\label{eq:TabPrev}
    \Tp_{ab} = \text{diag}(\rhop(t),-\pp(t) / (1 - kr^2),-\pp(t) r^2,-\pp(t) r^2\sin^2\theta)    
\end{equation}
with equation of state $\rhop = 3\pp$. The Einstein equations for this metric reduce to the Friedmann equations
\begin{align*}
    \Big{(}\frac{a'(t)}{a(t)}\Big{)}^2 
    &= \frac{\lambda}{3} - \frac{k}{a(t)^2} + \frac{\kappa}{3}\rho(t), \\
    \frac{a''(t)}{a(t)}
    &= \frac{\lambda}{3} - \frac12\kappa p(t) - \frac16\kappa\rho(t).
\end{align*}
The divergence free condition $\ddp_a\Tp^{ab} = 0$ can be integrated to yield $\rhop(t) = \mmp/\ap(t)^4$ where we call $\mmp$ the density parameter. When combined with the Friedmann equations and written in terms of proper time $\tau$ through integrating $\text{d}\tau = \text{d}t/\ap(t)$ yields
\begin{equation}\label{eq:Friedmann}
    a'(\tau)^2 = \kappa\frac{m}{3} - ka(t)^2 + \frac{\lambda}{3}a(t)^4.
\end{equation}

Throughout we use conventions of Penrose and Rindler \cite{penrose1984spinors}. The paper is organized as follows: Sec. \ref{sec:Tod} summarizes Tod's class of solutions and presents a new class that satisfies Penrose's original assumptions, Sec. \ref{sec:Newman} extends Newman's CCC solution to arbitrary spatial curvature $k$ and cosmological constant $\lambdap$ and clears up inconsistencies that were commented upon, Sec. \ref{sec:Nurowski} puts forward a minor correction and generalization of a small side remark of Nurowski's, Sec. \ref{sec:EMT} presents new generalized expressions for the energy-momentum tensor induced in the future aeon, Sec. \ref{sec:CF} discusses which of the many choices of fixing the conformal factor are satisfied by these solutions and Sec. \ref{sec:Discussion} summarizes the work.

\section{A class of radiation fluid solutions}\label{sec:Tod}
\subsection{Tod's class of solutions}\label{sec:TodsSolns}
Tod \cite{tod2015equations} imposes an FLRW universe on each side of the crossover in the bandage region given by Eq. \eqref{eq:FLRW} and does not make any restrictions on the spatial curvature nor the scalar curvature of the bridging metric. No explicit solutions $\ap$ or $\af$ to the corresponding Friedmann equation are calculated. Rather, he chooses $\op = c_1\ap$, which satisfies Eq. \eqref{eq:RicciTrafo} with 
\begin{equation}\label{eq:Todlambda}
    R = 4\lambda=6kc_1^2,
\end{equation}
and further specifies $c_1 = (\lambdap/\mmp)^{1/4}$ so that the future aeon is also governed by the same Friedmann equation as the previous aeon with the identifications $\mmf = \mmp$ and $\lambdaf = \lambdap$. One immediately sees that Penrose's original assumptions of $\lambdap = \lambda = \lambdaf$ are violated by this choice of $c_1$. 

\subsection{A new class of solutions}\label{sec:OurSolns}
If one wanted to adhere to Penrose's $\lambdap=\lambda$ assumption, then an alternate choice $\displaystyle c_1 = \sqrt{\frac{2\lambdap}{3k}}$ can be used. This yields another class of solutions to CCC where the density parameter and cosmological constant are no longer the same in each aeon, but rather satisfy
\begin{equation}
    \mmf = \frac{9k^2}{4\lambdap},\qquad
    \lambdaf = \frac{4\lambdap^2\mmp}{9k^2}.
\end{equation}
An important property of this class of solutions is that $\mmf = \mmp$ and $\lambdaf=\lambdap$ iff the additional relationship
\begin{equation}\label{eq:sameparams}
    \mmp\lambdap = \frac94k^2
\end{equation}
holds. If this is taken to be satisfied then our new class of solutions and Tod's are the same. Thus, if one wants to satisfy that the cosmological constant is the same in all three spaces while maintaing $\lambdap=\lambda=\lambdaf$, an additional restriction between the density parameter and cosmological constant must be satisfied. Further, it is easy to see that if one slightly violates this relationship while enforcing $\lambdap=\lambda$, either the cosmological constant will iterate to zero and the density parameter will iterate to positive infinity, or vice versa, showing that this solution is an unstable bifurcation. This can be seen explicity by the recurrence relations
\begin{equation}
    \lambda_{i+1} = \frac{4\lambda_i^2m_i}{9k^2}, \qquad
    m_{i+1} = \frac{9k^2}{4\lambda_i},
\end{equation}
where $\lambda_i$ and $m_i$ are the cosmological constant and density parameter associated to aeon $i$. These can be written in terms of an initial choice of $\lambda_0$ and $m_0$ as
\begin{equation}
    \lambda_i = \Big{(}\frac49m_0\Big{)}^{i-1}\lambda_0^i, \qquad
    m_i = \Big{(}\frac49\lambda_0\Big{)}^{1-i}m_0^{2-i}, \qquad
    i\in\mathbb{N}.
\end{equation}
These sequences cannot both converge unless $m\lambda = (4/9)k^2$ in which case we have $\mmf = \mmp$ and $\lambdaf = \lambdap$ and so these parameters are the same in every aeon.


\section{A perturbative approach}\label{sec:Newman}
\subsection{Extending Newman's solution}
Newman considers nearly the same setting as Tod, namely an FLRW universe on each side of the crossover in the bandage region given by Eq. \eqref{eq:FLRW}. However, he additionally assumes spatial flatness $k=0$, fixes $\lambdap=3$ and chooses the bridging metric to have zero scalar curvature $R=\lambda=0$. Unlike Tod, he obtains explicit scale factors as solutions of the Friedmann equation in the form of power series expansions. Importantly, these choices again differ from Penrose's original assumptions of $\Rp = R$ and $k=1$. In fact, Newman finds an inconsistency with the physically-motivated expansions around $\mathcal{C}$ laid out in \cite{gurzadyan2013ccc}. He guesses, backed by private communication with Tod, that this is generated from the inconsistency in the choice of $k$. Here, we extend Newman's calculations by allowing arbitrary $k,\lambdap$ and $R$ and show explicitly that this inconsistency actually arises by Newman's incorrect use of an expansion from \cite{gurzadyan2013ccc}. We will use a different notation to \cite{newman2014fundamental} to allow for a connection to be made with the other solutions presented here.

First, from the reciprocal hypothesis and that each aeon has a metric of the form Eq. \eqref{eq:FLRW}, we obtain that
\begin{equation}
    \op^4 = \Big{(}\frac{\ap}{\af}\Big{)}^2
    \Leftrightarrow
    \op^2 = -\frac{\ap}{\af},
\end{equation}
We solve the Friedmann equation \eqref{eq:Friedmann} by perform perturbations in the neighbourhood of $\mathcal{C}$ w.r.t. conformal time $\tau$. Defining $\Gamma := \taup - \taup_\infty = \tauf$ so that $\Gamma$ vanishes on $\mathcal{C}$ coming from either aeon, we find that 
\begin{align}
    \ap 
    &= -\frac{1}{\sqrt{\Lambdap}}\Gamma^{-1} 
    - \frac{k}{6\sqrt{\Lambdap}}\Gamma
    + \frac{1}{360\sqrt{\Lambdap}}(36\Mp\Lambdap - 7k^2)\Gamma^3 \\&+ \frac{1}{15120\sqrt{\Lambdap}}(108k\Mp\Lambdap - 31k^3)\Gamma^5,\nonumber \\
    \af
    &= \sqrt{\Mf}\Gamma 
    - \frac{k}{6}\sqrt{\Mf}\Gamma^3
    + \frac{1}{120}\sqrt{\Mf}(12\Mf\Lambdaf + k^2)\Gamma^5,
\end{align}
where we have defined $\Lambda := \lambda/3$ and $M := m/3$ along with the analogous definitions with a hat and check. These yield the expressions
\begin{align}
    \op^2 
    &= \frac{1}{\sqrt{\Lambdap\Mf}}\Gamma^{-2}
    + \frac{k}{3\sqrt{\Lambdap\Mf}}
    + \frac{2k^2 - 3(\Mp\Lambdap + \Mf\Lambdaf)}{30\sqrt{\Mf\Lambdap}}\Gamma^2, \\
    \op
    &= -\frac{1}{(\Mf\Lambdap)^{1/4}}\Gamma^{-1} 
    - \frac{k}{6(\Mf\Lambdap)^{1/4}}\Gamma
    - \frac{7k^2 - 18(\Mp\Lambdap + \Mf\Lambdaf)}{360(\Mf\Lambdap)^{1/4}}\Gamma^3.\label{eq:Op}
\end{align}
Noting that $g_{ab} = -\ap\af\tilde{g}_{ab}$, it is useful to also compute $a^2 := -\ap\af$, which takes the form
\begin{equation}\label{eq:asq}
    a^2 = \sqrt{\frac{\Mf}{\Lambdap}} - \frac{1}{10}\Mp\sqrt{\Mf\Lambdap}\Gamma^4.
\end{equation}
Then the Yamabe equation \eqref{eq:RicciTrafo}, written with respect to the bridging metric's Levi-Civita connection, can be written with $\Gamma=x^0$ as
\begin{equation}
    a^2\op'' + 2aa'\op'+\frac23\lambda a^4\op = \frac23\lambdap a^4\op^3,
\end{equation}
where a prime denotes $\text{d}/\text{d}\Gamma$. Using Eqs \eqref{eq:asq} and \eqref{eq:Op}, we find that the equation can be solved at all orders by the choice
\begin{equation}\label{eq:lambda}
    \lambda = \frac{3k}{2}\sqrt{\frac{\Lambdap}{\Mf}}
    \Leftrightarrow
    R = 6k\sqrt{\frac{\Lambdap}{\Mp}}.
\end{equation}
Note that, unlike in \cite{newman2014fundamental}, at this stage we have not needed to make any assumptions about how the previous aeon's parameters relate to the future aeon's. This arises from allowing $k\neq0$. Further, it is clear that Eq. \eqref{eq:lambda} is the same as Tod's in Eq. \eqref{eq:Todlambda} with his choice $c_1 = (\lambdap/\mmp)^{1/4}$.


\subsection{Penrose's conditions near $\mathcal{C}$}
The philosophy of CCC requires that rest-mass $\mu$ plays no role in the neighbourhood of $\mathcal{C}$. Recall that this is given by
\begin{equation}
    \mu = \frac{4}{\kappa}[2\lambda\op^2 - \lambdaf - \lambdap\op^4 + 3(1 - \op^2)^2\Pi_a\Pi^a], \tag{\ref{eq:trace}}
\end{equation}
where
\begin{equation}
    \Pi_a := \frac{\text{d}\op}{\op^2 - 1} = \frac{\text{d}\of}{1 - \of^2}.
\end{equation}
The quantity $\Pi_a$ is built from the conformal factor $\op$ (or $\of$), is unaffected by use of the reciprocal hypothesis $\op\rightarrow-\of^{-1}$ and is smooth through $\mathcal{C}$.

Noting that $\of$ can act as a time coordinate in the neighbourhood of $\mathcal{C}$, Penrose has given conditions that constrain the degrees of freedom in the Yamabe equation \eqref{eq:RicciTrafo} by considering the power series expansion 
\begin{equation}
    \Pi^a\Pi_a = A + B\op + C \op^2 + \ldots
\end{equation}
around $\mathcal{C}$ \cite{gurzadyan2013ccc} for the special case $\lambdap=\lambda=\lambdaf=3$. In this case, Eq. \eqref{eq:trace} reduces to
\begin{equation}
    \mu = 3\frac{4}{\kappa}(\op^2 - 1)^2(\Pi^a\Pi_a - 1).
\end{equation}
Thus, to supress the appearance of rest-mass in the future aeon, in accordance with the principles of CCC, he makes the power series expansion ansatz
\begin{equation}
    \Pi^a\Pi_a = 1 + Q\of^2 + O(\of^3).
\end{equation}
The linear term is taken to vanish while the quadratic term $Q$ is taken as a universal constant relating to a Higgs-type mechanism. He dubs the above the \emph{supressed rest-mass hypothesis}. This then imposes two conditions on the Yamabe equation, but does not yet fix a unique solution. Assuming the above expansion, one can obtain other useful expansions which contain $Q$. These are
\begin{equation}
    \nabla^a\of\nabla_a\of = 1 + (Q-2)\of^2 + O(\of^3), \qquad
    \nabla^a\Pi_a = 2Q\of + O(\of^2).
\end{equation}
Newman finds that these expansions are inconsistent with his results, which he guesses spawn from his different choice $\lambda=0\neq\lambdap$. We find this is in fact not the reason. We present generalized expansions with no assumptions on $\lambdap$ and $\lambda$ to show this. They become
\begin{align}
    \Pi^a\Pi_a 
    &= \frac{\lambdap}{3} + Q\of^2 + O(\of^3), \label{eq:P1} \\
    \nabla^a\of\nabla_a\of
    &= \frac{\lambdap}{3} + (Q - \frac23\lambdap)\of^2 + O(\of^3), \label{eq:P2}\\
    \nabla^a\Pi_a 
    &= [\frac23(\lambda - \lambdap) + 2Q]\of + O(\of^2).\label{eq:P3}
\end{align}
When $\lambda=0,\lambda=3$ we see that Eq. \eqref{eq:P3} reduces to
\begin{equation}
    \nabla^a\Pi_a 
    = 2(Q - 1)\of + O(\of^2),
\end{equation}
At this point we note that it was not correct for Newman to compare his expansion using the specializations $\lambdap=3,\lambda=k=0$ to $\nabla^a\Pi_a = 2Q\of + O(\of^2)$, as this expression is only valid when $\lambda=\lambdap=3$ via Eq. \eqref{eq:P3}. Thus, the inconsistency found when expanding $\nabla^a\Pi_a$ was \emph{not} caused by his choice of $k=0$.

When using the solution to the Yamabe equation \eqref{eq:lambda} in these three expressions for $Q$, we find they are consistent in general and yield
\begin{equation}
    Q = \frac23\lambdap - k\sqrt{\frac{\lambdap}{\mmf}}.
\end{equation}
One sees that in Newman's case with $k=0$ and $\lambdap=3$ then $Q=2$ as expected. Thus, any choice of $\lambda$ or $k$ will be consistent with these expansions, no further constraints are produced and there are no inconsistencies.

\section{Nurowski's solution}\label{sec:Nurowski}
As a minor side remark that is not at all the main result of his paper, Nurowski puts forward a solution very similar to Tod but with an explicit scale factor and for $k=1$ \cite{nurowski2021poincare}. We find an error in this solution and present here a correction which is consistent when compared to Tod's solution and our extension of Newman's, as well as being generalized to arbitrary $k$. In doing so, we use the same conventions as we have been using throughout, which differ from \cite{nurowski2021poincare}.

Writing down the Einstein equations for the previous aeon with metric and energy-momentum tensor given by Eqs \eqref{eq:FLRW} and \eqref{eq:TabPrev} respectively but without fixing an equation of state yields
\begin{equation}
    \lambdap(1 + \omegap)\ap(t)^2 
    - (1 + 3\omegap)(k + \ap'(t)^2)
    - 2\ap(t)\ap''(t) 
    = 0,
\end{equation}
where we have now chosen $\omegap := \pp / \rhop$ as a constant. The corresponding equation in the future aeon related through $\gf_{ab} = \op^{-4}(t)\gp_{ab}$ is
\begin{gather}\label{eq:Omegappfuture}
    -(1 + 3\omegaf)(k + \ap'(t)^2)\of(t)^4
    + 2(\ap(t)\of(t)^3)[2(2 + 3\omegaf)\ap'(t)\of'(t)
    - \of(t)\ap''(t)] \nonumber \\
    + \ap(t)^2[\lambdaf(1 + \omegaf)
    + 4\op(t)^2(-(2 + 3\omegaf)\op'(t)^2
    + \op(t)\op''(t))]
    = 0,
\end{gather}
where we have now chosen $\omegaf := \pf / \rhof$ as a constant also. We can here make the choice $\op(t) = c_1\ap(t)$, and eliminating $\ap''(t)$ from the above two equations gives
\begin{gather}\label{eq:Together}
    -\lambdaf(1 + \omegaf) 
    + kc_1^2(2 + 3\omegap)\op(t)^2 \nonumber \\
    +\op(t)^2[-\lambdap(1 + \omegap)\op(t)^2
    + (2 + 3\omegap)\op'(t)^2
    + 3\omegaf(kc_1^2 + \op'(t)^2]
    = 0.
\end{gather}
This can be solved directly, but following the procedure in \cite{nurowski2021poincare} we can take another derivative, use Eq. \eqref{eq:Omegappfuture} to eliminate $\op''(t)$ and assuming $\op'(t) \neq 0$ obtain
\begin{gather}
    -\lambdaf(1 + \omegaf)[2 + 3(\omegap + \omegaf] \nonumber \\
    + \op(t)^2\Big{(}6c_1^2 - 4\lambdap\op(t)^2 + 6\op'(t)^2
    + 3\omegaf(5 + 3\omegaf)(kc_1^2 + \op'(t)^2) \nonumber \\
    + \omegap[9kc_1^2(1 + \omegaf)
    -4\lambdap\op(t)^2
    + 9(1 + \omegaf)\op'(t)^2]\Big{)}
    = 0.
\end{gather}
Finally, replacing $\op'(t)$ in the above using Eq. \eqref{eq:Together} yields the simple equation
\begin{equation}
    \lambdaf(1 - 3\omegap)(1 + \omegaf)
    - \lambdap(1 + \omegap)(1 - 3\omegaf)\op(t)^4
    = 0.
\end{equation}
Neglecting having a constant $\op(t)$ implies we must satisfy
\begin{equation}
    \lambdaf(1 - 3\omegap)(1 + \omegaf) = 0, \qquad
    \lambdap(1 + \omegap)(1 - 3\omegaf) = 0.
\end{equation}
The solutions are
\begin{gather}
    \omegap = \omegaf = -1, \\
    \omegap = \omegaf = \frac13, \label{eq:omegasoln} \\
    \lambdap = \lambdaf = 0, \\
    \lambdap = 0, \qquad
    \omegap = \frac13 \text{ or } 
    \omegaf = -1, \\
    \lambdaf = 0, \qquad
    \omegap = -1 \text{ or } 
    \omegaf = 1/3.
\end{gather}
We choose Eq. \eqref{eq:omegasoln} which corresponds to $\Tp_{ab}$ and $\Tf_{ab}$ being trace-free. In this case, Eq. \eqref{eq:Together} is integrated to yield a solution which replaces and generalizes Theorem (2.4) of \cite{nurowski2021poincare}
\begin{gather}
    \op(t)^2 
    = \frac{1}{48\lambdap}e^{-2(t + c_2)\sqrt{\lambdap/3}}
    \Big{[}9\Big{(}e^{2c_2\sqrt{\lambdap/3}}
    + 4kc_1^2e^{2t\sqrt{\lambdap/3}}\Big{)}^2
    - 64\lambdap\lambdaf e^{4t\sqrt{\lambdap/3}}\Big{]},
\end{gather}
with integration constant $c_2$.

At this stage we have essentially the same result as Tod, but now with arbitrary $\lambdap$ and $\lambdaf$ and importantly, an explicit analytic expression for $\op(t)$ or equivalently $\ap(t)$. One can check that the scalar curvature of the bridging metric with this conformal factor is indeed $6kc_1^2$ in line with Tod. Thus, the bridging metric has constant scalar curvature and the Yamabe equation \eqref{eq:RicciTrafo} is satisfied.

\section{The energy-momentum tensor in the future aeon}\label{sec:EMT}
The full energy momentum tensor $\check{U}_{ab}$ of the future aeon, namely the one that satisfies the Einstein equation with associated metric $\gf_{ab}$, has been fixed as a perfect radiation fluid. However, Penrose has a prescription for this in terms of three parts
\begin{equation}
    \check{U}_{ab} = \Tf_{ab} + \Tf_{ab}[\op] + \check{W}_{ab},
\end{equation}
where $\Tf_{ab}$ is the contribution of $\Tp_{ab}$, $\Tf_{ab}[\op]$ is the contribution of the scalar field $\op$ and $\hat{W}_{ab}$ has yet to obtain a physical interpretation but generically includes a trace-term that Penrose associates with dark matter. 

Given $\Pi_a := (\nabla_a\op) / (\op^2 - 1)$ and $\Phi_{ab} = (-1/2)[R_{ab} - (1/4)Rg_{ab}]$ is the trace-free Ricci tensor of the bridging metric, the individual parts have the following general expressions 
\begin{gather}
    \Tf_{ab} = \op^4\Tp_{ab}, \\
    \Tf_{ab}[\op] = 2\op^2\Phi_{ab}, \\
    \check{W}_{ab} = \Big{[}3(1 - \op^2)^2 \Pi^c\Pi_c - \lambdaf + 2\lambda \op^2 - \lambdap\op^4\Big{]}\gf_{ab} \nonumber \\
    + 2(1 - \op^2- \op^4)\Phi_{ab} + 2(\op^{-1} + \op^3)(\nabla_a\nabla_b\op - \frac14\nabla_c\nabla^c\op g_{ab}) \nonumber \\
    - 4\op^2(\nabla_a\op\nabla_b\op - \frac14\nabla_c\op\nabla^c\op g_{ab}), \\
    \check{W}_{ab}\gf^{ab} = \mu = 4[2\lambda\op^2 - \lambdaf - \lambdap\op^4 + 3(1 - \op^2)^2\Pi_a\Pi^a]. \label{eq:trace}
\end{gather}
These expressions are checked against the CCC solution of the previous section. The three pieces are easily defined in terms of the future aeon $\gf^{ab}$ as
\begin{gather}
    \Tf^{ab} = \frac13\Rf\of(t)^8\delta^a_0\delta^b_0 - \frac{1}{12}\Rf\of(t)^4\gf^{ab}, \\
    \Tf^{ab}[\op] = \frac13R\of(t)^8\delta^a_0\delta^b_0 - \frac{1}{12}R\op(t)^4\gf^{ab}, \\
    \hat{W}^{ab} = \frac13(\Rp - \Rf - R)\op(t)^8\delta^a_0\delta^b_0 - \frac{1}{12}(\Rp - \Rf - R)\op(t)^4\gf^{ab},
\end{gather}
where it is immediate that they are all trace-free. Commenting upon the individual physical interpretation of these is not easy, other than how they are sourced by quantities of the previous aeon. The new-improved energy momentum tensor $\Tf_{ab}[\Omega]$ is in fact the conserved energy tensor for solutions of the conformally invariant wave equation \cite{penrose1984spinors2}. Finding a corresponding Lagrangian could shed some light, but we do not pursue this. Of particular note is that $\hat{W}_{ab}$ if $k\neq0$ and $\lambdaf \neq \lambdap$ to avoid $c_1=0$, then one can choose
\begin{equation}
    c_1^2 = \frac{2}{3k}(\lambdap - \lambdaf),
\end{equation}
so that $\hat{W}_{ab}$ vanishes.

\section{Penrose's candidates for constraining $\op$}\label{sec:CF}
There are multiple choices in how to fix the remaining freedom in the Yamabe equation for the conformal factor \cite{penrose2010cycles,tod2021conformal,tod2015equations}. Although these and essentially one general FLRW CCC model have currently been put forward, it remains to see which conformal factor choices are satisfied by these models. Here, we take the choice choice of $\op = c_1\ap$ that has been used in the solutions presented above. To be as general as possible, we present the results of Sec. \ref{sec:Nurowski}.

The choices to fix the conformal factor first presented in \cite{penrose2010cycles} are
\begin{subequations}
    \begin{align}
        N^aN^b\Phi_{ab} &= O(\of^3), \label{eq:CFCond1} \\
        N^aN^b\nabla_aN_b &= O(\of^2), \label{eq:CFCond2} \\
        \nabla_a\Pi^a &= O(\of), \label{eq:CFCond3} \\
        3\Pi_a\Pi^a - \lambdap &= O(\of^3), \label{eq:CFCond4}
    \end{align}
\end{subequations}
where $N_a = \nabla_a\of$. These expansions for the given CCC model are
\begin{subequations}
    \begin{align}
        N^aN^b\Phi_{ab} 
        &= \frac14c_1^2k\lambdap
        - \frac34c_1^4k^2\of^2
        +O(\of^4), \\
        N^aN^b\nabla_aN_b
        &= -\frac13c_1^2k\lambdap\of
        + O(\of^3), \\
        \nabla_a\Pi^a
        &= (\frac23\lambdap - kc_1^2)\of + O(\of^3), \\
        3\Pi^a\Pi_a - \lambdap
        &= (2\lambdap - 3kc_1^2)\of^2
        + O(\of^4).
    \end{align}
\end{subequations}
It is clear that Eq. \eqref{eq:CFCond3} is the only condition satisfied in general by this CCC solution. The only way that Eqs. \eqref{eq:CFCond1} and \eqref{eq:CFCond2} can be satisfied is if $k=0$ which has the consequence that $\lambda=0$ and Penrose's $\lambdap=\lambda$ hypothesis is violated. Eq. \eqref{eq:CFCond4} can be satisfied by choosing $c_1^2 = 2\lambdap/(3k)$ provided $k\neq0$. This is also the choice that implies $\lambdap=\lambda$ and satisfies Penrose's hypothesis delayed rest-mass hypothesis with $Q=0$. 

It is then clear that for the CCC solution considered, only Eq. \eqref{eq:CFCond4} gives a reasonable condition to fix the conformal factor by way of restricting $c_1$. Of note is that Tod \cite{tod2015equations} and Nurowski \cite{nurowski2021poincare} give their own, different conditions by requiring the metric at the crossover to have a particular form coming from a Starobinksy expansion in the previous aeon \cite{starobinsky37isotropization}, but we do not discuss these here.


\section{Discussion}\label{sec:Discussion}
In this paper we have presented a new class of solutions to the CCC paradigm that describe consequtive FLRW universes with a radiation fluid. This class is very nearly the same as that presented by Tod \cite{tod2015equations} but where the original assumptions of Penrose \cite{penrose2010cycles} are satisfied. These solutions have the consequence that the energy density $\rho(t) = m / a(t)^4$ and the cosmological constant $\lambda$ will blow up and converge to zero respectively, or vice-versa, unless an additional relationship between $m$ and $\lambda$ is satisfied. This relationship, namely $m = 9k^2/4\lambda$, implies that these parameters will remain constant throughout all aeons. It is shown that this bifurcation is unstable. In the case that this relationship is satisfied, Tod's class and this new class of solutions are identical.

Newman \cite{newman2014fundamental} also looked at successive radiation fluid FLRW universes, but explicitly solved the Friedmann equations with series expansions for the scale factors and specialized to $\lambda=3$ and $k=0$ in each aeon. He commented on an inconsistency in certain expansions of Penrose \cite{gurzadyan2013ccc} around the crossover that he put down to Penrose's alternate choice of $k=1$. We generalized Newman's calculations to arbitrary $\lambda$ and $k$ and show that his choice of $k=0$ was not the cause of this inconsistency, but rather it was his use of an expansion in \cite{gurzadyan2013ccc} that was only valid when $\lambdap = \lambda = 3$. It was shown that the expansions given in \cite{gurzadyan2013ccc} are perfectly consistent for this CCC solution for arbitrary $k$ and cosmological constants so long as the Yamabe equation \eqref{eq:RicciTrafo} is satisfied.

Nurowski \cite{nurowski2021poincare} has the only other solution to CCC in the current literature, which again makes use of consecutive FLRW universes with a radiation fluid. He proposes a conformal factor that joins two aeons around the crossover which yields an explicit form of the scale factor. This generalizes Tod's solution in that each aeon may have a different cosmological constant, and generalizes Newman's solution in that the expressions are explicit. We found that this conformal factor, given as a minor side-remark in, was incorrect and presented a new one that was justified by comparing to the other solutions in the respective limits. This yields the most general, explicit class of solutions to the CCC paradigm that currently exists in the literature. 

The energy-momentum tensor of the future aeon was broken up into three individually conserved pieces following \cite{penrose2010cycles} and new general expressions were given for these in terms of the bridging metric's connection. These were evaluated with our general solution and commented upon.

A major missing piece in the CCC paradigm is a physically motivated method of fixing the conformal factor uniquely. All solutions so far impose a constant scalar curvature for the bridging metric, which implies the Yamabe equation \eqref{eq:RicciTrafo} must be satisfied. The uniqueness of the conformal factor then reduces to how to uniquely solve this equation. While Tod and Nurowski have proposed fixing this based upon utilizing a Starobinksy expansion around the crossover, Penrose proposes several other candidates for doing this \cite{penrose2010cycles}. These candidates are tested using the CCC model derived here and surprisingly, only one can be satisfied, namely the condition
\begin{equation*}
    3\Pi_a\Pi^a - \lambdap = O(\of^3),
\end{equation*}
which we take as a leading contender to fix the conformal factor for more general classes of solutions of the CCC paradigm.

\backmatter

\bmhead{Acknowledgments}
The authors wish to thank N. Bishop, J. Frauendiener, A. Goodenbour and P. Tod for useful discussions.

\section*{Declarations}
Data sharing not applicable to this article as no datasets were generated or analysed during the current study.

\bibliography{papers}

\end{document}